\documentstyle[10pt,epsf]{article}

\title{When do finite sample effects significantly affect
	entropy estimates ?} 

\author{\begin{tabular}{c}  
T. Dudok de Wit
\\
\\
{\small Centre de Physique Th\'eorique, CNRS and 
	Universit\'e de Provence, Marseille, France}\\
\\
{\small accepted in Eur. Phys. J. B}
\end{tabular}}

\date{}

\begin{document} 

\maketitle

\begin{abstract}
An expression is proposed for determining the error
made by neglecting finite sample effects in entropy
estimates.  It is based
on the Ansatz that the ranked distribution of
probabilities tends to follow a Zipf scaling.
\end{abstract}

\vspace*{10mm}

%%%%%%%%%% section 1 %%%%%%%%%%%%%%%%%%%%%%%%%%%%%%%%%%%%%%%%%%%%%

\section{Introduction}

The growing interest in complexity measures 
and symbolic dynamics \cite{Beck93,Badii97} 
has brought to the forefront various problems
related to the estimation of entropic quantities from
finite sequences \cite{Schuermann96}. Such estimates
are known to suffer from a bias, which 
prevents quantities such as the
metric entropy from being meaningfully estimated. 
The purpose of this letter is to provide an analytical
expression for this bias,in order in order to test for
finite sample effects in entropy estimates.

\medskip

Consider the general case of a string of $N$ symbols
$\{i_1 i_2 \cdots i_N\}$, each of which belongs to a
finite alphabet ${\mathcal A}$. The average informational content
of substrings of length $d$ taken from this sequence is
expressed by the Shannon entropy
\cite{Blahut87}
\begin{equation}
\label{eq:entropy}
	H_d = - \sum_{i_1,\ldots,i_d \in {\mathcal A}}
	\mu \left([i_1 i_2 \cdots i_d]\right) \log
	\mu \left([i_1 i_2 \cdots i_d]\right) \ ,
\end{equation}
where $\mu$ is the natural invariant measure with respect
to the shift. 
Of particular interest is the
block or dynamical Shannon entropy $h_d = H_{d+1}-H_d$ 
from which one gets
the measure-theoretic entropy of the system
\begin{equation}
	h(\mu) = \lim_{d \rightarrow \infty} h_d  \ ,
\end{equation}
a quantity that is intimately related to the
Kolmogorov-Sina\"{\i} entropy in case the string represents 
the output of a shift dynamical system.

The main problem lies in the estimation of the empirical
measure $\mu$ from a finite string of symbols. Direct
box counting yields
\begin{equation}
\label{eq:boxcount}
	\mu \left([i_1 i_2 \cdots i_d]\right) \approx 
	\frac{\# [i_1 i_2 \cdots i_d]}{N-d+1} \ ,
\end{equation}
where $\# [i_1 i_2 \cdots i_d]$ is the occurrence
frequency of the block $i_1 i_2 \cdots i_d$ in the string.
It is well known that statistical fluctuations
in the sample on average lead to a systematic 
underestimation of the
entropy. This problem becomes particularly acute as the word
size increases for a given string length $N$. 
Since this deviation can easily be
mistaken for the
signature of a finite memory process, it is of prime 
importance to determine whether its origin is physical
or not.

Several authors have already addressed the problem of 
making corrections to empirical entropy estimates
\cite{Schuermann96,Herzel88,Grassberger88,Schmitt93}; their
expressions are valid as long as the occurrence 
frequencies of the observed words are large compared to
one. While this may hold for relatively short words,
it breaks down for long ones, making it difficult for
a small correction to be used as a safe indication for 
a small deviation. Our objective is to derive a more reliable 
(although less accurate) expression of the deviation, 
to be used as a warning signal against the
onset of finite sample effects.

As a first guess one could 
require the sample to be long enough for each word
to have a chance to appear. This gives 
$N \gg N_{\textrm{\scriptsize symb}}^d$, where
$N_{\textrm{\scriptsize symb}}$ 
is the cardinality of the alphabet. This
criterion, however, is generally found to be too
conservative because it does not take into account
the grammar, i.e. the rules that cause some words to be 
forbidden or less frequent than others.

%%%%%%%%%% section 2 %%%%%%%%%%%%%%%%%%%%%%%%%%%%%%%%%%%%%%%%%%%%%%%%%%%%

\section{The Zipf-ordered distribution}

To derive our expression, we first rank the words
according to their frequency of occurrence: let
$n_{k=1}$ denote the frequency of occurrence
of the most probable word, $n_{k=2}$ of the next most
probable one etc. Multiple instances of the same
frequency get consecutive ranks. This monotonically decreasing
distribution is called Zipf-ordered.

The Asymptotic Equipartition Property introduced by
Shannon \cite{Blahut87} states that the ensemble of words of
length $d$ can be divided into two subsets. The first one
consists of ``typical words'' that occur frequently and roughly
have the same probability of occurrence. The other subset
is made of ``rare words'' that belong to the tail
of the distribution. According to the
Shannon-Breiman-MacMillan theorem, the entropy is 
related to the typical words in the limit where
$N \rightarrow \infty$; the contribution of rare words
progressively disappears as $N$ increases.
In some sense this observation
justifies the procedure to be described below.

It was noted by Pareto \cite{Pareto97}, Zipf \cite{Zipf49}
and others, 
and later interpreted by Mandelbrot \cite{Mandelbrot97}
that the tail of the Zipf-ordered distribution
$n_k$ tends to follow a universal scaling law
\begin{equation}
	n_k = \alpha k^{-\gamma} \ , \ \ \gamma > 0 \ .
\end{equation}
which is found with astonishing reproducibility in
economics, social sciences, physics etc. 
\cite{Mandelbrot97}. As
shown in \cite{Gunther96,Troll98}, many different systems 
give rise to Zipf laws, whose ubiquity is thought to
be essentially a consequence of the ranking procedure.

The physical meaning of Zipf's law is still
an unsettled question, although it does not seem to 
reflect any particular self-orga\-niza\-tion 
(see for example \cite{Miller58,Li98}).
We just mention that a slow decay
is an indication for a ``rich vocabulary'', in the
sense that rare words occur relatively often.

The key point is that the empirical Zipf-ordered
distribution has a cutoff at some finite
value $k=N_{\textrm{\scriptsize max}}$ because of 
the finite length of the symbol string. For the same reason, the
occurrence frequencies are necessarily quantized. 
Our main hypothesis is that the true
distribution extends beyond $N_{\textrm{\scriptsize max}}$, 
up to the lexicon size $K \ge N_{\textrm{\scriptsize max}}$, 
following Zipf's law with the same exponent $\gamma$.
This Ansatz has already been
suggested as a way to estimate entropies from
long words \cite{Poschel95}.

%%%%%%%%%% section 3 %%%%%%%%%%%%%%%%%%%%%%%%%%%%%%%%%%%%%%%%%%%%%%%%%%%%

\section{Estimating the bias}

Let $\hat{H}$ be the Shannon entropy computed from the
empirical distribution (using eqs.~\ref{eq:entropy} and
\ref{eq:boxcount}) and $H$ the entropy one would obtain from
a non truncated distribution, in which the frequencies are not 
quantized anymore and extend beyond $N_{\textrm{\scriptsize max}}$
following Zipf's law.
\begin{eqnarray}
\label{eq:twoentropies}
	\hat{H} &=& - \sum_{k=1}^{N_{\textrm{\tiny max}}} 
		\frac{n_k}{\sum_{k=1}^{N_{\textrm{\tiny max}}} n_k} \log
		\frac{n_k}{\sum_{k=1}^{N_{\textrm{\tiny max}}} n_k} \ , \\
	H &=& - \sum_{k=1}^K 
		\frac{n_k}{\sum_{k=1}^K n_k} \log
		\frac{n_k}{\sum_{k=1}^K n_k} \ . \nonumber
\end{eqnarray}
The truncation has two counteracting effects. It
changes the renormalization of the occurrence
frequencies and causes some of the least frequent words
to be omitted.

The difference $\delta$ between the two entropy
estimates  
\begin{equation}
	\delta = H - \hat{H} \ .
\end{equation}
is what we call the bias, to be used as a measure of 
the deviation resulting from finite sample effects. 
We shall assume that $N_{\textrm{\scriptsize max}} \gg 1$,
which is equivalent 
to saying that the distribution must have a sufficiently
long tail for a power law to make sense.

It is natural to define a small parameter 
$0 \le \varepsilon \ll 1$, which goes to zero for 
a non truncated distribution
\begin{equation}
	\varepsilon = \frac{1}{N}\sum_{k=N_{\textrm{\tiny max}}+1}^K n_k
		\ .
\end{equation}
Remember that $N = \sum_{k=1}^K n_k$ \cite{Comment1}.

Now, assuming that Zipf's law persists for
$k>N_{\textrm{\scriptsize max}}$, we have
\begin{equation}
	\varepsilon
	= \frac{1}{N}\!\sum_{k=N_{\textrm{\tiny max}}+1}^K \alpha k^{-\gamma}
	= \frac{\alpha}{N} \big( 
	\zeta(\gamma,N_{\textrm{\scriptsize max}}+1) 
	- \zeta(\gamma,K+1) \big) \ , 
\end{equation}
where $\zeta(\gamma,m)$ is the Hurwitz or generalized
Riemann zeta function. For $k > \gamma$, the following
approximation holds \cite{Spanier87}
\begin{equation}
	 \zeta(\gamma,m)
	= \frac{m^{1-\gamma}}{\gamma-1}
		- \frac{m^{-\gamma}}{2}
		+ \frac{m^{-\gamma-1}}{12}  \ .
\end{equation}
Since $K, N_{\textrm{\scriptsize max}} \gg 1$, we may write
\begin{equation}
	\varepsilon
	= \frac{\alpha}{N(\gamma-1)}
		\left({N_{\textrm{\scriptsize max}}^{1-\gamma} 
		- K^{1-\gamma}}\right) \ .
\end{equation}
The value of $\alpha$ remains to be determined.
To do so, we note that the least frequent 
words in the Zipf-ordered distribution occur once or a few
times only. One may therefore reasonably set 
$n_{k=N_{\textrm{\tiny max}}} \approx 1$,
giving $\alpha \approx N_{\textrm{\scriptsize max}}^{\gamma}$. 

The bias $\delta$ can now be expanded in
powers of $\varepsilon$. Keeping terms of order
${\mathcal O}(\varepsilon)$ only, we have
\begin{equation}
	\delta = -\varepsilon \hat{H} + (1+\varepsilon)
		\left(\varepsilon - 
		\sum_{k=N_{\textrm{\tiny max}}+1}^K 
		\frac{n_k}{N} \log \frac{n_k}{N} \right)  \ .
\end{equation}
For the conditions stated before, the sum can be 
approximated by
\begin{equation}
	\sum_{k=N_{\textrm{\tiny max}}+1}^K  
	\frac{n_k}{N}  \log  \frac{n_k}{N}
	= - \varepsilon  \left( \log N - \gamma 
	\log \frac{N_{\textrm{\scriptsize max}}}{K}
	\right) \ ,
\end{equation}
finally giving the result of interest
\begin{eqnarray}
\label{eq:delta}
	\delta &=& \varepsilon  \left( 1+\log N - \hat{H} 
	- \gamma  \log \frac{N_{\textrm{\scriptsize max}}}{K} \right) \\
	\varepsilon &\approx& \frac{N_{\textrm{\scriptsize max}}}{N (\gamma-1)}
	\left( 1 - \left( \frac{N_{\textrm{\scriptsize max}}}{K} \right)^{\gamma-1} \right) 
	\ . \nonumber
\end{eqnarray}
Notice that the true entropy is always underestimated;
furthermore $\varepsilon$ is continuous at $\gamma = 1$
\cite{Comment2}.
Most of the variation comes from the small
parameter $\varepsilon$, whose expression
reveals two different effects :
\begin{enumerate}

\item the ratio $N_{\textrm{\scriptsize max}}/N$ 
reflects the uncertainty of the frequency estimates.

\item the scaling index $\gamma$, whose value is usually
between 0.5 to 1.5, is indicative of the lacunarity
of the word distribution. In the case of a shift
dynamical system, $\gamma$ reveals how unevenly the rare orbits 
fill the phase space.

\end{enumerate}
For the sake of comparison, the first order
approximation for finite sample effects
derived in \cite{Herzel88,Grassberger88} is
\begin{equation}
	\delta = \frac{N_{\textrm{\scriptsize max}}}{2 N} \ .
\end{equation}

We conclude from eq.~\ref{eq:delta} that the bias is not just 
related to statistical fluctuations
in the empirical occurrence frequency, 
but is also caused by the omission of words that
are asymptotically rare.
If the true distribution of the
ranked words were exponential or ultimately ended
with an exponential tail, then our criterion would 
be too conservative but still reliable as such.

The following procedure is proposed for detecting
the maximum word length for which entropies can be
meaningfully estimated :  compute Zipf-ordered distributions 
for increasing word-lengths $d$. For each length, estimate
the bias $\delta$ by least-squares
fitting a power law to the tail of the observed distribution. 
As soon as this bias exceeds a given threshold 
(say 10\% of $\hat{H}$), then entropies computed from longer 
words are likely to be significantly corrupted by finite 
sample effects.

Equation \ref{eq:delta} supposes that the maximum lexicon size
$K$ is known a priori, which is seldom the case. This is not
a serious handicap, however, since the value of $K$ has relatively
little impact on the bias; a rough approximation such as
$K=N_{\textrm{\scriptsize symb}}^d$ may do well.

%%%%%%%%%% examples %%%%%%%%%%%%%%%%%%%%%%%%%%%%%%%%%%%%%%%%%%%%%%%%%%%%

\section{Two examples}

To briefly illustrate the results, we now consider two
examples. The first one is based on a Bernouilli
process, whose entropy and Zipf-ordered distribution
can be calculated analytically.
The string of symbols is drawn from a two letter alphabet, one
with probability $\lambda$ and the other with probability
$1-\lambda$. The block
entropy of this process is independent of the word length and
equals
\begin{equation}
	h = - \lambda \log \lambda - (1-\lambda) \log (1-\lambda) 
		\ .
\end{equation}

Figure 1 compares the true block entropy with estimates drawn
from a sample of length $N=2000$ with $\lambda=0.15$. 
The departure of the
empirical estimate from the true one is evident. Without
knowledge of the true entropy, however, it is very
difficult to tell whether the decrease of the entropy is an
artifact or just the signature of a short-time memory.

%%%%%%%% figure 1 %%%%%%%%%%%%%%%%%%%%%%%%%%%%%%%%%%%%%%%%%%%%%%%

\begin{figure}[!htb] 
\centerline{\epsfbox{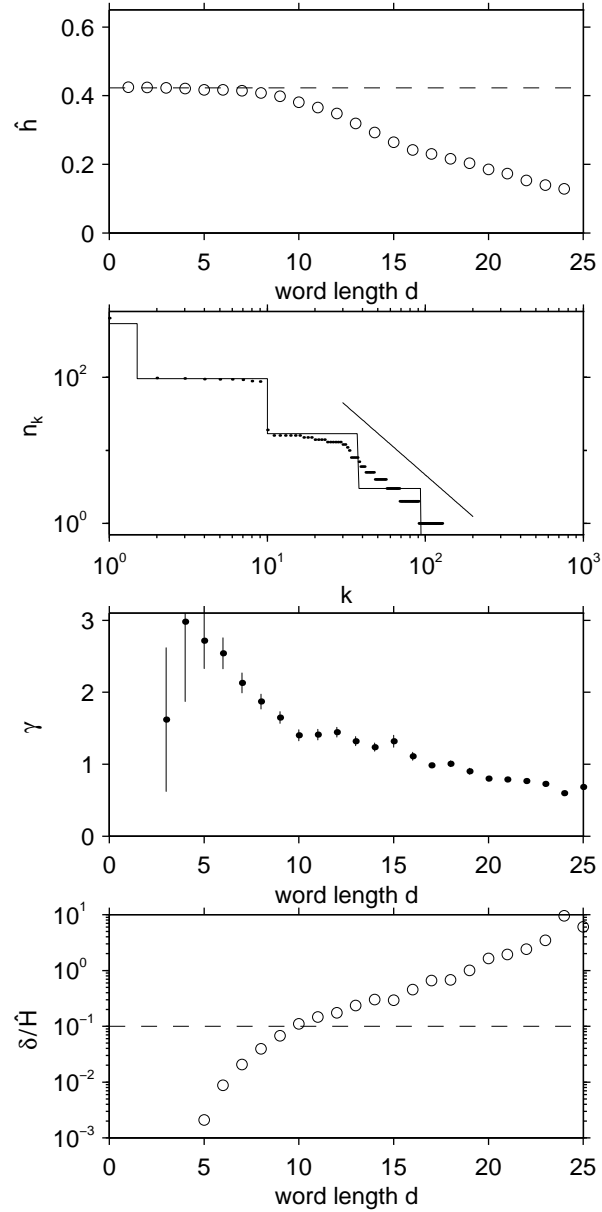}}
\caption{Analysis of a Bernouilli sequence, with
$N=2000$ and $\lambda=0.15$. From top to bottom: (1) the 
empirical block entropy and the true one (dashed),
(2) the true (line) and observed (dots) Zipf-ordered distributions 
for words of length $d=8$; (3) the scaling exponent $\gamma$ 
obtained by fitting the tail of the Zipf-ordered distribution 
(error bars represent $\pm 1$ standard deviation resulting
from the least-squares fit), 
(4) the bias $\delta$. In this case, entropies cannot
be reliably estimated for word lengths beyond $d=9$.
Block entropies are normalized to 
$\log N_{\textrm{\scriptsize symb}}$, so that the
maximum possible value is 1.}
\label{fig1}
\end{figure}

The second panel displays the true and the empirical Zipf-ordered
distributions as obtained for words of length $d=9$. Zipf's law
clearly holds for words whose rank exceeds about 30. 
After this, the scaling exponent $\gamma$ is estimated, see the
third panel. The decrease of this exponent with the word 
length $d$ suggests that the contribution
of the rare words becomes increasingly important. 
Finally, the bias $\delta$,
which is shown in the fourth panel, suggests that the
onset of a significant bias occurs around $d=8$; this value
is indeed in agreement with the results of the first panel.

The validity of the bias estimate was tested
on various examples and was found to be reliable,
provided that $N_{\textrm{\scriptsize max}} \gg 1$. 

\begin{sloppypar}
In the second example, we consider a sequence of 
$N=10^4$ symbols generated by the logistic map 
$x_{i+1} = \lambda x_i(1-x_i)$ in a chaotic regime
with $\lambda = 3.8$. The (generating) partition 
${\mathcal P} = \{[0, 0.5[, [0.5, 1]\}$ gives
us a two-letter alphabet.
\end{sloppypar}

%%%%%%%% figure 2 %%%%%%%%%%%%%%%%%%%%%%%%%%%%%%%%%%%%%%%%%%%%%%%

\begin{figure}[!htb] 
\centerline{\epsfbox{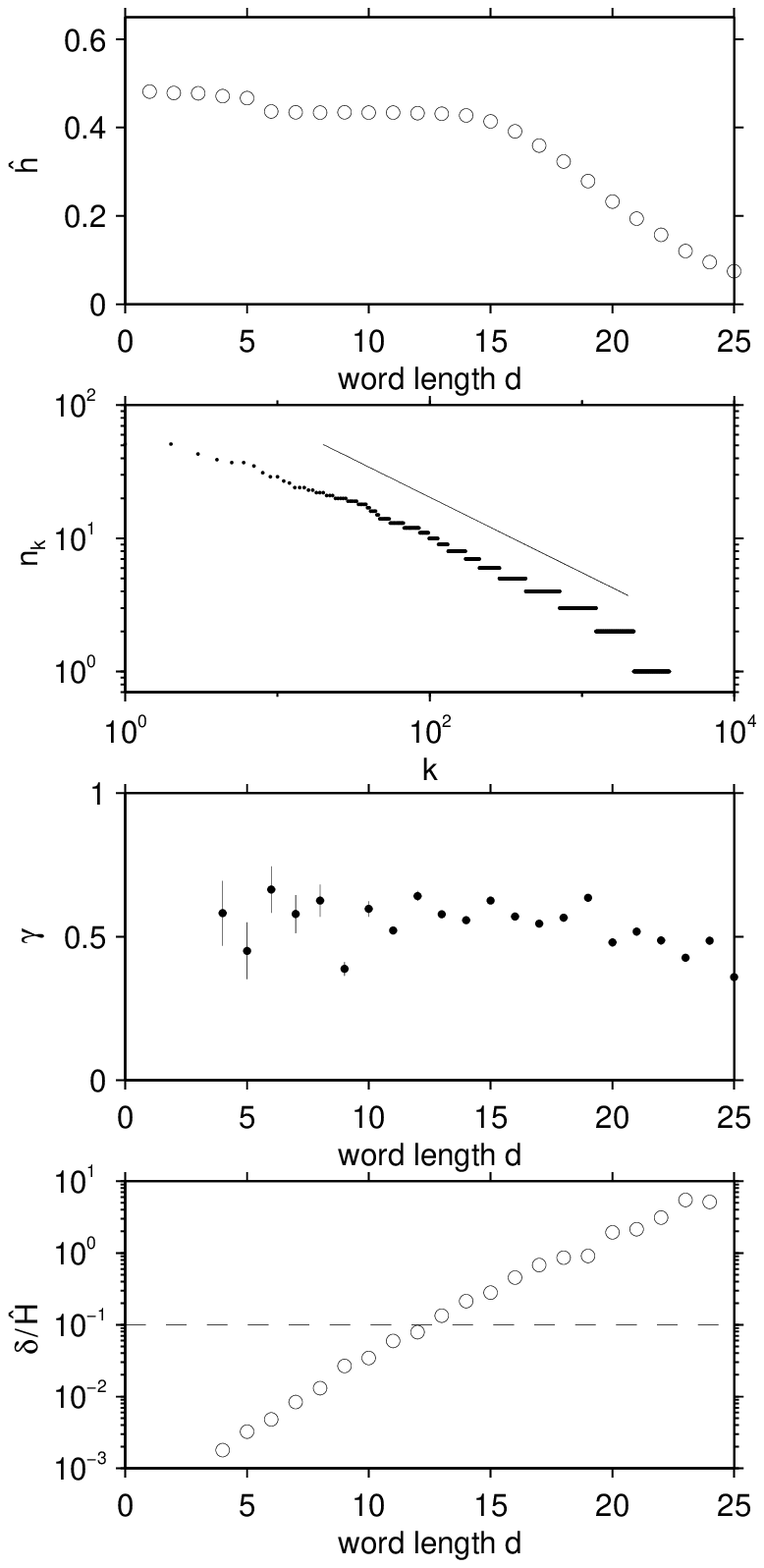}}
\caption{Analysis of a logistic map sequence, with the same
legend as the previous figure; the string length
is $N=10^4$. The second panel shows
a Zipf-ordered distribution for $d=18$. The largest 
word size for which the relative bias is smaller
than 10$\%$, is $d=12$.}
\label{fig2}
\end{figure}

Figure 2 again shows that the block entropy decreases
above a certain word length. In contrast to the previous
example, the measured scaling exponent $\gamma$ is small and
almost constant, regardless of the word length. We believe 
this to be a consequence of the intricate structure
of the self-similar attractor. 
This low value of $\gamma$ already suggests
that rare words should bring a significant contribution
to the entropy. The bias $\delta$ finally suggests 
stopping at $d=12$.

%%%%%%%%%% conclusion %%%%%%%%%%%%%%%%%%%%%%%%%%%%%%%%%%%%%%%%%%%%%

\clearpage

\section{Conclusion}

Summarizing, we have derived a simple expression
(eq. \ref{eq:delta}) 
for detecting the onset of finite
sample size effects in entropy estimates. It is based
on the empirical evidence that rank-ordered distribution
of words tend to follow Zipf's law. The criterion
reveals that rare events can significantly
bias the empirical entropy estimate.

%%%%%%%%%% references %%%%%%%%%%%%%%%%%%%%%%%%%%%%%%%%%%%%%%%%%%%%%


\begin{thebibliography}{}
\bibitem{Comment1}
To be exact $N-d+1=\sum_{k=1}^K n_k$, but 
we use the fact that $d \ll N$.
\bibitem{Comment2}
The term $(K/N_{\textrm{\scriptsize max}})^{1-\gamma}$ 
can be large when $K \gg N_{\textrm{\scriptsize max}}$ 
and $\gamma<1$ but this divergence
becomes effective long after the 
maximum word size has been exceeded; it is therefore not
a matter of concern here.
\end{thebibliography}
\end{document}